\documentclass[12pt,preprint]{aastex}

\def\Msun{M_\odot}

\def\kmsmpc{\rm{~km~s^{-1}~Mpc^{-1}}}

\slugcomment{Submitted to The Astrophysical Journal}
\lefthead{Agustsson \& Brainerd}
\righthead{Locations of Satellite Galaxies}

\begin{document}

\title{The Locations of Satellite Galaxies in a $\Lambda$CDM Universe}

\author{Ingolfur Agustsson \& Tereasa G. Brainerd}
\affil{Boston University, Institute for Astrophysical Research,
725 Commonwealth Ave., Boston, MA 02215}

\email{ingolfur@bu.edu, brainerd@bu.edu}

\begin{abstract}
We compute the locations of satellite
galaxies with respect to their hosts using the
$\Lambda$CDM GIF simulation.
If the major axes of the hosts' images are perfectly aligned
with the major axes of their projected mass, the satellites
are located preferentially close to the hosts' major axes.  In this case,
the degree of anisotropy in the satellite locations is a good tracer of the
flattening of the hosts' halos.  If all hosts have
luminous circular disks, the symmetry axes of the projected mass and light
are not perfectly aligned, and the locations of the
satellites depend upon how
the hosts' disks are placed within their halos.  If the disk angular momentum
vectors are aligned with the major axes of the halos, the satellites
show a pronounced ``Holmberg effect''.  If the disk angular momentum vectors are
aligned with the intermediate axes of the local large scale structure, the
distribution of satellite locations is essentially isotropic.  If the disk angular
momentum vectors are aligned with either the minor axes or
with the net angular momentum vectors of the halos, the satellites are
distributied anisotropically about their hosts,
with a preference for being found nearby the
hosts' major axes.  This agrees well with the observation that satellite
galaxies in the Sloan Digital Sky Survey tend to be found nearby the major axes
of their hosts, and suggests that the mass and light of SDSS host galaxies must be
fairly well aligned in projection on the sky.

\end{abstract}

\keywords{dark matter --- galaxies: dwarf --- galaxies: fundamental parameters ---
galaxies: halos --- galaxies: structure }

\section{Introduction}

The standard cold dark matter (CDM) scenario predicts that large, bright 
galaxies reside within mildly--flattened halos that accrete mass
preferentially along filaments.  Recent work on weak galaxy--galaxy lensing
(e.g., Guzik \& Seljak 2002; Hoekstra et al.\ 2004; Kleinheinrich et al.\ 2004) 
and the kinematics of satellite galaxies (e.g., Prada et al.\ 2002; 
Brainerd 2004) suggests
that the spherically--averaged potentials of large field galaxies are in
quite good agreement with the predictions of CDM (i.e., the Navarro, Frenk \&
White, NFW, profile; Navarro, Frenk \& White 1995, 1996, 1997).  
Within the virial radius, $r_{200}$, the median projected
ellipticity of CDM galaxy halos is $\epsilon_{\rm halo} \sim 0.3$.
On scales $r << r_{200}$, the
effects of gas cooling will make the halos somewhat rounder that this, but
on scales $r \sim r_{200}$ the shapes of the halos are not greatly affected
by the baryons (e.g., Kazantzidis et al.\ 2004.)  
In order to fully test the CDM paradigm
one would ideally like to compare this prediction of flattened CDM halos to
the observed shapes of the dark halos in our universe.  Direct
constraints on the actual shapes of dark galaxy halos 
are, however,
much more difficult to obtain than are constraints on the spherically--averaged
halo potentials (see, e.g., the review by Sackett 1999). 

Hoekstra et al.\ (2004) found that their galaxy--galaxy lensing
signal was consistent with the halos of the lens galaxies being flattened
to the degree expected from CDM: $\epsilon_{\rm halo} = 0.33^{+0.07}_{-0.009}$.
This is an exciting result, but is a bit controversial for two reasons.  First,
structures larger than galaxies (i.e., nearby groups or clusters of galaxies)
may contribute an external shear that could affect the inferred flattening
of galaxy--mass halos.  Second, Hoekstra et al.\ (2004) made two
simplifying assumptions about their lens galaxies: (1) mass and light are
perfectly aligned in projection on the sky, and (2) the ellipticities of the 
dark halos of the lens
galaxies are related to the ellipticities of the observed
images through
a constant multiplicative factor, $\epsilon_{\rm halo} = f \epsilon_{\rm lens}$,
where $\epsilon_{\rm lens}$ is the ellipticity of the light emitted by the lens
galaxies.

Here we investigate another possible indicator of the shapes of dark galaxy
halos: the location
of small, faint satellite galaxies with respect to the symmetry axes of large,
bright ``host'' galaxies.  This approach is complimentary to galaxy--galaxy
lensing since, in both cases, an ensemble average over many galaxies
is necessary to detect a signal.  
In the case of galaxy--galaxy lensing, the weak lensing  shear is too
small to be detected convincingly from a single lens galaxy and, hence, thousands
of lenses are needed.  In the case of the locations of
satellite galaxies, standard host--satellite selection algorithms generally
yield only one or two satellites per host in the large redshift surveys.
Therefore, a convincing
measurement of the locations of satellites relative to their hosts requires a
large number of objects.

Our study is motivated by a recent finding that the satellites of 
isolated host galaxies in the Sloan Digital Sky Survey are distributed
anisotropically about their hosts (Brainerd 2005).  In particular, on scales
less than a few hundred kiloparsecs, the SDSS satellites are located
preferentially close to
the major axes of their hosts.  This is the exact
opposite of the so--called ``Holmberg effect'' (e.g., Holmberg 1969; Lynden--Bell 1982;
Majewski 1994; Zartisky et al.\ 1997), in which satellite galaxies are found
preferentially close to the minor axes of their hosts.  It could
be argued that the disagreement between these early studies and that of
Brainerd (2005) is merely the result of small number statistics in the early
samples of hosts and satellites.  However, Sales \& Lambas (2004) analyzed the
location angles of satellite galaxies in the Two Degree Field Galaxy Redshift
Survey and, using a sample size similar to that of Brainerd (2005),
Sales \& Lambas (2004) concluded that the majority of the 2dFGRS satellites were 
distributed isotropically
about their hosts.  In a very small, restricted subsample of their
data, however, Sales \& Lambas (2004) found weak evidence for the 2dFGRS
satellites to be located preferentially close to the minor axes of the hosts
(i.e., evidence for the Holmberg effect).

This disagreement between Brainerd (2005) and Sales \& Lambas (2004) has
fueled controversy over whether or not satellite galaxies have a preferred
location relative to their hosts.  A recent reanalysis of the 2dFGRS
data by Sales \& Lambas has, however, revealed an error in the host position angles
in the original data
and when the error is corrected the satellites of the 2dFGRS hosts show
the same anisotropy that was found by Brainerd (2005): a preference
for clustering near the major axes of the host galaxies (Sales \& Lambas 2006, in
preparation).

On the theoretical side, previous numerical work leads us to expect that
satellite galaxies in CDM universes will not be spherically--distributed
around their host galaxies.  Knebe et al.\ (2004) found that the
orbits of satellites
of primary galaxies in cluster environments were located preferentially
within a cone of opening angle 40$^\circ$.
The structure of CDM halos is
largely independent of the halo mass scale (e.g., Moore et al.\  1999), so
this suggests that the satellites of
isolated host galaxies in CDM models
ought to be roughly aligned with the major 
axes of the host halos.  Further,  recent numerical work by Libeskind et al.\ (2005)
and Zentner et al.\ (2005) on the luminous satellites
of Milky Way--type halos has shown that the satellites tend to lie
in highly--flattened structures that are essentially embedded in the principle
planes of the host halos (i.e., the plane defined by the major and intermediate
moments of the inertia tensor). 
Our study complements these investigations by using a much larger and, hence
more statistically significant, sample of objects.

To explore the possibility that the location of satellite galaxies relative
to their hosts may serve as a tracer of the dark mass distribution around 
host galaxies, we use the $\Lambda$CDM GIF simulation (Kauffmann et al.\ 1999)
to pose the following questions:

\begin{itemize}
\item Are satellite galaxies in a $\Lambda$CDM universe distributed isotropically or
anisotropically about their hosts?
\item How does the distribution of satellites 
compare to the distribution
of dark mass surrounding the hosts?
\item If the light emitted by the hosts comes from a disk, how does the orientation
of the disk within the host's halo affect the inferred satellite distribution?
\end{itemize}

\noindent
Throughout, we analyze the simulation in the same way in which an observer
would analyze a combined imaging and redshift survey.  That is, we work
in terms of the locations of objects and the shapes of halos as 
seen in
projection on the sky, and we select host and satellite galaxies from
the simulation using the
same types of algorithms that are used to select hosts and satellites from
observational data.

\noindent
The outline of the paper is as follows.  In \S2 we discuss the GIF simulation
and the way in which host galaxies and their satellites are selected.  In 
\S3 we make a simple assumption that the mass and light of
host galaxies are perfectly aligned in projection on the sky and
we compute the location angles of the satellite galaxies 
with respect to the major axes of
the projected host halos.  In \S4 we model the luminous regions
of the host galaxies as circular disks and we embed the disks within the
halos according to various prescriptions.  We then compute the location angles
of the satellite galaxies with respect to the major axes of the projected
host disks.  A summary and discussion of our results is given in 
\S5.

\section{Hosts and Satellites in the $\Lambda$CDM GIF Simulation}

The GIF simulations are a suite of CDM simulations which combine 
adaptive P$^3$M N--body techniques
with semi--analytic galaxy formation.  The inclusion of semi--analytic galaxy
formation eliminates the ``overmerging problem'' in which galaxies within halos
that merge to form larger structures (i.e., groups and clusters) quickly
lose their identities as individual objects.  The location of a luminous
galaxy in the GIF 
simulations is initially identified with the most bound particle in a given
halo.  When two or more halos merge, the luminous galaxies within the halos
maintain their separate identities with the exception that the luminous
galaxies may ultimately
merge on a time scale that is set by dynamical friction.
It is these luminous galaxies whose locations are associated with 
individual particles that we use to identify
hosts and satellites within the simulation and, hence, the satellite population
which we investigate does not suffer from an artificial overmerging problem.  
For a complete discussion of the way in which luminous galaxies are allowed to
merge in the GIF simulations, see \S 4.4 and \S 4.8 of Kauffmann et al.\ (1999).

Since it seems that our universe is consistent with having cosmological
parameters of $\Omega_{m0} = 0.3$, $\Lambda_0 = 0.7$, and
$H_0 = 70 \kmsmpc$, we use only the $\Lambda$CDM GIF simulation for our analysis.
The $\Lambda$CDM GIF
simulation has these particular cosmological parameters,
a co--moving box size of 201.7~Mpc, a mass per particle
of $2\times 10^{10} \Msun$, and a softening length of 28.6~kpc.  The simulation
can be downloaded from the GIF project website, 
\url{http://www.mpa-garching.mpg.de/GIF}, for a wide range of redshifts.  Here
we use primarily the present--epoch data since the large redshift surveys
(i.e., SDSS and 2dFGRS)
that have been used to investigate the observed location of satellite galaxies
relative to their hosts are restricted to fairly low 
redshift.  

We specifically use the $z=0$ $\Lambda$CDM GIF galaxy catalog in which the magnitudes
of the galaxies are given in the SDSS band passes (i.e., the file called
galsl\_sdss.cat\_1178).  In addition,
we use the dark matter particle file called compd4001.1178.
The luminous galaxies in the simulation have known properties such as stellar
mass, luminosity, and color, as well as known locations and peculiar velocities.
Dark matter halos which surround the luminous galaxies must necessarily be 
identified from the mass particles via a halo finding routine.  A file called 
halos.propl\_1178 that contains dark halo information is provided by the 
GIF group, but we choose not to use this file.  The information in
halos.propl\_1178 is based upon halos that were found using a friends--of--friends
algorithm, which is known to often link into one single object
two or more nearby halos that have
distinct, identifiable centers.  
Instead of friends--of--friends halos, then, we simply use the 
particle file compd4001.1178 to define
dark matter halos as the mass contained within spheres of radius $r_{200}$, where
$r_{200}$ is the radius inside which the mean interior mass density is equal
to 200 times the critical mass density.  This definition of the virial radius
and virial mass is consistent with the formalism adopted by
Navarro, Frenk \& White (1995, 1996, 1997).

For our analysis, we require the dark matter halos of the
host galaxies to have a sufficiently large number
of particles for a reasonable measurement of the halo shape
to be made.  Therefore, we restrict our analysis to host galaxies with
dark matter halos that contain 100 or more particles within $r_{200}$
(i.e., the minimum halo mass for our host galaxies is $2\times 10^{12} \Msun$).
We make no such minimum mass restriction on the satellite galaxies.
The virial radii and the centers of the host galaxy halos are computed in
3 dimensions using a standard iterative scheme.  The halo is initially
assumed to be
centered on the location
of the luminous galaxy.  The virial radius is then defined to be the
radius of a sphere of particles, centered on the luminous galaxy,
for which the mean interior mass density is equal to 200 times the critical
density.  The center of mass of the initial sphere of particles
is computed, and this is then used to define a new center 
from which a new sphere of particles
is obtained and yet another center of mass is computed.  The process is repeated
until convergence is reached.  Convergence occurs within only a few iterations
and, in projection on the sky, the location of the 
particle that represents the position of the luminous galaxy is typically
offset by less than half a smoothing length from 
center of mass of the galaxy's halo.
  
Host--satellite systems are selected by rotating the simulation
randomly and then projecting the simulation along the line of
sight.  The resulting line of sight velocities, coordinates on the ``sky'', and the
apparent magnitudes of the GIF galaxies are then used as direct substitutes
for the type of data that would be available in a large redshift survey.
Results below are obtained from 100 random rotations of
the simulation.  Following Brainerd (2005), three different methods
are used to select isolated hosts and their satellites via a
combination of line of sight velocity difference,
$|dv|$, projected radius from the host, $r_p$,
and apparent magnitude difference, $\Delta m \equiv m_1 - m_2$,
where $m_1 > m_2$ .
Throughout we take the coordinates and peculiar velocities of the GIF galaxies
to be those of the individual particles that flag the presence of a
luminous galaxy in the simulation.
In Sample 1, the apparent magnitude difference between a host galaxy
and any other galaxy that lies within
$r_p < 700$~kpc and 
$|dv| < 1000$~km~s$^{-1}$ must be $\Delta m \ge 1.0$.
Satellites of Sample 1 hosts must fall within
$r_p < 500$~kpc
and $|dv| < 500$~km~s$^{-1}$ and have $\Delta m \ge 2.0$.  In Sample 2,
the apparent magnitude difference between a host galaxy and
any other galaxy that
lies within $r_p < 2.86$~Mpc and 
$|dv| <  1000$~km~s$^{-1}$ must be $\Delta m \ge 0.75$.   Satellites of
Sample 2
hosts must lie within
$r_p < 500$~kpc and 
$|dv| < 1000$~km~s$^{-1}$ and have $\Delta m \ge 1.5$.  
In Sample 3, the magnitude difference between a host galaxy and 
any other galaxy that lies within 
$r_p < 500$~kpc and 
$|dv| < 1000$~km~s$^{-1}$ must be $\Delta m \ge 2.25$.  
Further, the magnitude difference between a host galaxy in Sample 3
and any other galaxy that lies within
$r_p < 1$~Mpc and
$|dv| < 1000$~km~s$^{-1}$ must be $\Delta m \ge 0.75$.  
Satellites of Sample 3 hosts
must lie 
within $r_p < 500$~kpc and
$|dv| < 500$~km~s$^{-1}$, and have $\Delta m \ge 2.25$.
These criteria select only very isolated hosts and
their satellites, and it is worth noting that both the Milky Way and M31 would
be rejected as host galaxies under these restrictions.

In order to eliminate a small number of systems that pass the above tests
but which are, in reality, more likely to
be representative of cluster environments rather than isolated host--satellite
systems, we impose two further restrictions: (1) the sum total of the luminosities
of the satellites of a given host must be less than the luminosity of
the host, and (2) the total number of satellites of a given host must be
less than 9.  In addition, 
we restrict our analysis to hosts with luminosities in the
range $0.5 L^\ast_{b_J} \le L \le 5.5 L^\ast_{b_J}$ since Brainerd (2004)
found that outside this luminosity range the kinematics of the satellites
in the $\Lambda$CDM GIF simulation
were not consistent with a virialized population.  
This last criterion eliminates only a small number of possible hosts from the
analysis ($\sim 2$\% of the Sample 1 hosts, $\sim 7$\% of the Sample
2 hosts, and $\sim 3$\% of the Sample 3 hosts).
To compute the $b_J$ magnitudes
of the GIF galaxies, the SDSS magnitudes given in the file galsl\_sdss.cat\_1178
were converted using the photometric transformation of Norberg et al.\ (2002):
\begin{equation}
b_J = g' + 0.155 + 0.152(g' -  r') .
\label{transform}
\end{equation}
For the cosmological parameters used in the $\Lambda$CDM GIF
simulation, the absolute magnitude of an $L^\ast$ galaxy in the
$b_J$ band is
$M_{b_J}^\ast = -20.43 \pm 0.07$.

After all of the above selection criteria have been imposed, we find that 
on average the
individual rotations of the simulation yield
1786 hosts and 5752 satellites in
Sample 1, 317 hosts and 1208 satellites
in Sample 2, and 949 hosts and 2865 satellites in Sample 3.

A final important point is that, while the host galaxies in the GIF
simulation have known luminosities, there are no actual {\it images} of
the galaxies in the simulation.  For our investigation, then, we need to
{\it define} an image for each of the host galaxies in order to determine
the locations of the satellites with respect to the symmetry axes of the
host images.  In all cases, we define the images of the host galaxies to be
ellipses on the sky; however, the orientations of the image
ellipses are obtained in a number of different ways.  To begin, we 
assume that the mass and the
light of the host galaxies are perfectly aligned in
projection on the sky.  In this case, the major
axis of the image of the host galaxy corresponds to the major axis of
the projected halo mass distribution.  Next, we assume that all host
galaxies are circular disks and we embed the disks within the hosts'
halos using various prescriptions.  In these cases, the major axis
of the image of a host galaxy is the major axis of the
host's circular disk
as seen in projection on the sky, and here the major axis of the
light is not necessarily aligned
with a symmetry axis of the projected halo mass.
We will justify our various choices for the host images
in the sections below.

\section{Alignment of Light and Mass in the Hosts} 

In this section we make a very simple assumption that the major axis
of the image of a host galaxy is perfectly aligned with the major 
axis of the projected mass distribution of its halo.  This assumption
can be partially justified by the argument that galaxies are relaxed
systems and, if the dark matter halos are substantially flattened,
then the most dynamically reasonable expectation is that mass and
light should be fairly well aligned.  Direct determinations of the 
degree of alignment of mass and light in observed galaxies are difficult
and rare, but in a study of strong lens galaxies Keeton et al.\ (1998) found
that the major axes of the mass and light in the lens galaxies were
aligned to within $10^\circ$ in projection on the sky.

To begin our analysis of the locations of satellite galaxies with respect
to their hosts, we use the mass contained within the spherical overdensity
region of radius $r_{200}$ around each host to compute the 
principle moments of inertia of the halo and, thus, the halo's equivalent ellipsoid.
This yields axis ratios $b/a$ and $c/a$ for each of the halos, where 
we define $a \ge b \ge c$.  A triaxiality parameter is then computed
for each of the host halos:
\begin{equation}
T = \frac{a^2 - b^2}{a^2 - c^2}.
\end{equation}
Here $T = 0$ indicates a purely oblate object and $T=1$ indicates a
purely prolate object.
To define the major axis of a host galaxy, we project the 3-dimensional
ellipsoid of its halo onto the sky.  This yields an ellipse with 
semi--major axis $\alpha$, semi--minor axis $\beta$, and ellipticity
$\epsilon_{\rm halo} = 1 - \beta / \alpha$.  
The major axis of the light emitted by the
host galaxy is then defined to be the major axis of the halo's
projected ellipsoid
and this is used as the symmetry axis relative to which the
location angles of the satellite galaxies are measured.

Figure~1 shows the distribution of host luminosities, the mass function of
the host halos, the distribution of the host halo triaxiality parameters,
the distribution of the number of satellites per host, the distribution of
apparent magnitude differences between the hosts and their satellites,
and the distribution of the ratio of stellar masses of the satellites and
hosts.
The median host luminosity is
$1.5 L^\ast_{b_J}$, $1.9 L^\ast_{b_J}$ and $1.8 L^\ast_{b_J}$ for Samples
1, 2 and 3, respectively.  The median virial mass of the host halos is
$3.5\times 10^{12} \Msun$, $4.1\times 10^{12} \Msun$ and $3.5 \times 10^{12}
\Msun$ for Samples 1, 2 and 3, respectively, and the median host halo
triaxiality is $T = 0.6$ for all three host--satellite samples.
The bottom panels of Figure~1 show that our algorithm for finding hosts
and satellites clearly selects satellites that are considerably smaller
and fainter than their hosts.

\begin{figure}
\centerline{
\scalebox{0.65}{%
\includegraphics{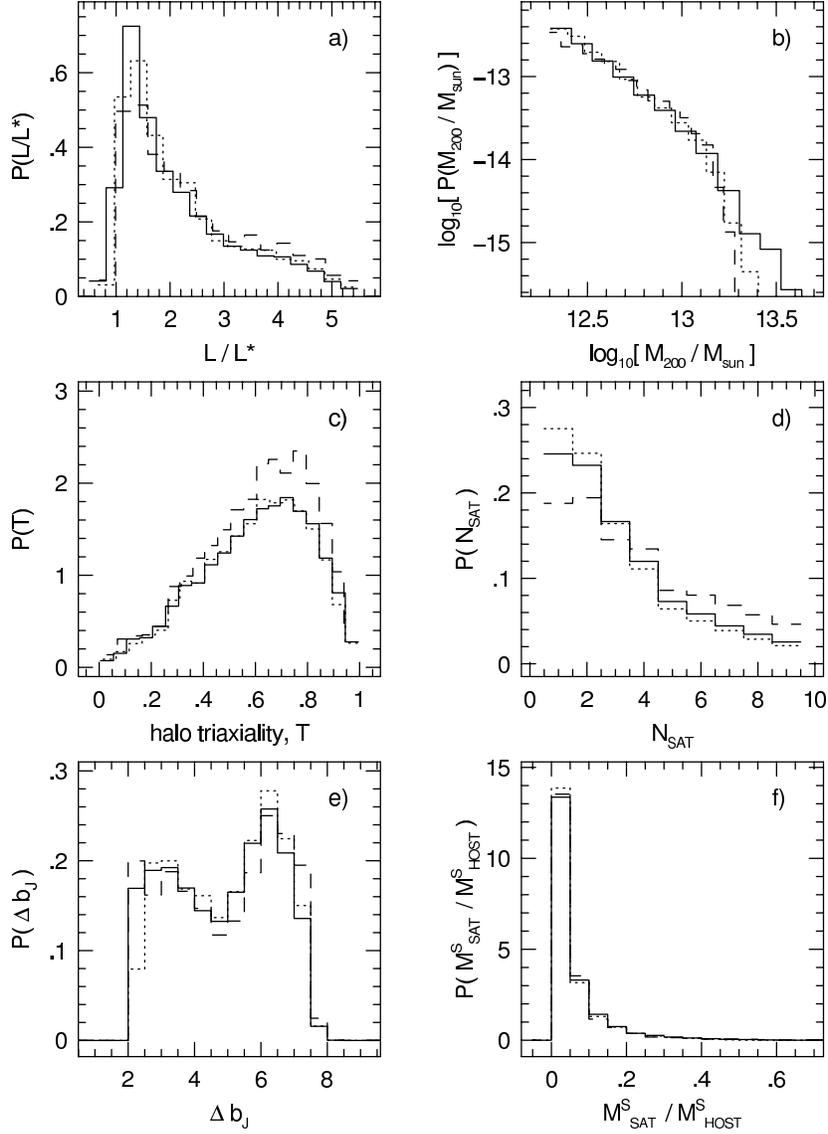}%
}
}
\caption{
a) Probability distribution of host luminosities, b) Host halo mass function,
c) Probability distribution of host halo triaxialities, d) Probability
distribution for the number of satellites
per host, e) Probability distribution for apparent magnitude differences
between hosts and satellites, $\Delta b_J \equiv b_J^{\rm sat} - b_J^{\rm host}$,
f) Probability distribution for the ratio of satellite stellar mass to
host stellar mass.
In all panels the different line types correspond to the different host--satellite
samples: Sample~1 (solid), Sample~2 (dashed), Sample~3 (dotted).
See text for host and satellite selection criteria.
}
\label{fig1}
\end{figure}

Since the host halos are not resolved particularly well (i.e., they contain
of order hundreds of particles, not thousands), 
for the remainder of this particular section
we restrict our analysis
to systems for which the projected ellipsoid  of the
host's halo has ellipticity 
$\epsilon_{\rm halo} > 0.2$.
This insures that the orientation of the major axis
of the host is well--determined.  When this ellipticity requirement is
imposed, the different rotations of the simulation yield an average of
1793 hosts and 5769 satellites in Sample 1, 320 hosts and 1209 satellites
in Sample 2, and 957 hosts and 2892 satellites in Sample 3.

For each of the host--satellite samples, we compute the location angles
of the satellites relative to the major axes of the hosts.  These are
simply polar angles, $\phi$, on the sky where $\phi = 0^{\circ}$ corresponds
to a satellite that is located along the direction of the host's major
axis and $\phi = 90^{\circ}$ corresponds to a satellite that is located
along the direction of the host's minor axis.  
Shown Figure~2 are the differential probability
distributions, $P(\phi)$, and the cumulative probability distributions,
$P(\phi \le \phi_{\rm max})$, for the location angles of the satellites
measured with respect to the major axes of their hosts.
Here the null hypothesis to which $P(\phi)$
and $P(\phi \le \phi_{\rm max})$ should be compared
is that the satellite galaxies are distributed spherically
about their hosts.  This would give rise to a circularly--symmetric
distribution of satellites on the
sky.  A rotation of
a spherically--symmetric
distribution through any combination of Euler angles always
gives rise to a
2--d distribution that is circularly--symmetric. Hence, any
deviation in the satellite distribution from pure circular symmetry
cannot be caused simply by projection and/or rotation effects and
must reflect an underlying non--spherical 3--d distribution of the satellites.

\begin{figure}
\centerline{
\scalebox{0.65}{%
\includegraphics{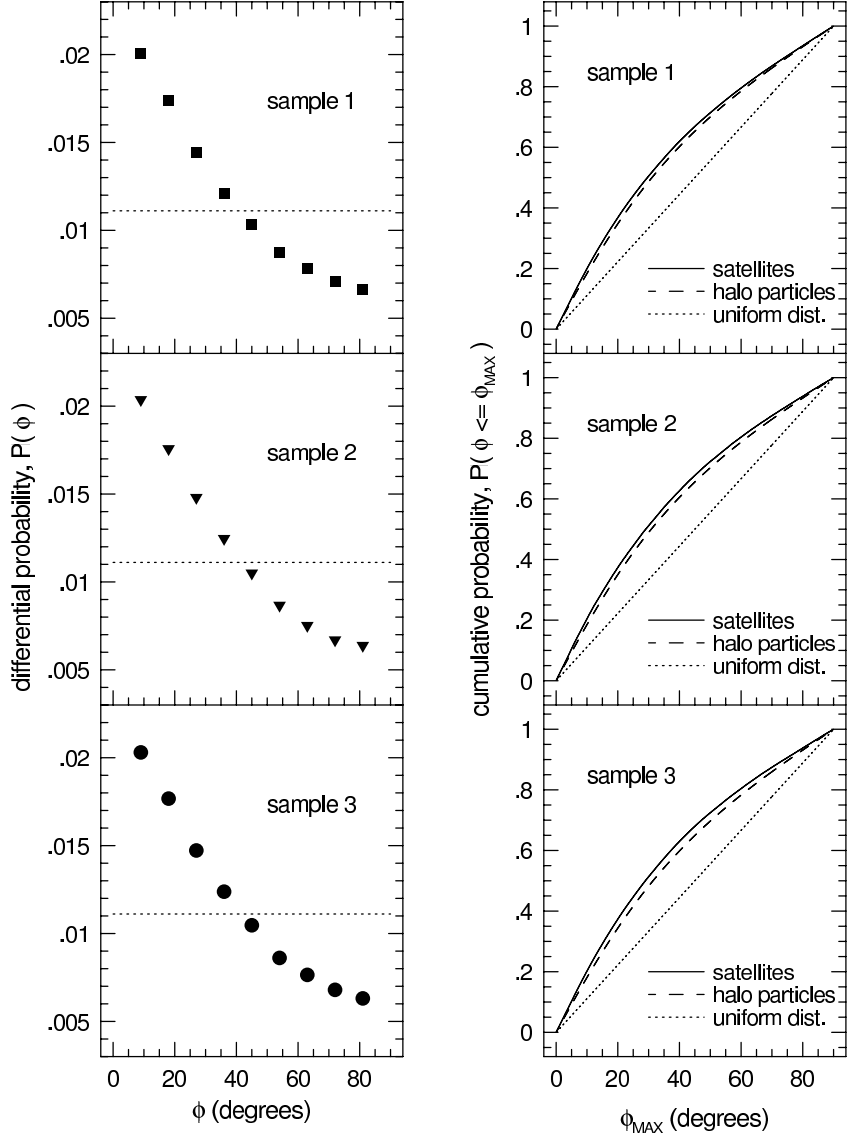}%
}
}
\vskip -1.0cm
\caption{
{\it Left panels:} Differential probability distribution function for the location
angles of the
satellites, measured with respect to the major axes of the projected halo mass.
Here $\phi = 0^\circ$ corresponds to alignment with the
projected halo major axis and $\phi = 90^\circ$ corresponds to alignment with
the projected halo minor axis.
Dotted line shows the expectation for a uniform (circularly--symmetric)
distribution of satellites.
{\it Right panels:} Cumulative probability distribution for the
location angles of the satellites (solid lines) and 
halo mass particles (dashed lines) with
respect to the major axes of the projected halo mass.  {\it Top
panels:} Sample~1. {\it Middle panels:} Sample~2.  
{\it Bottom panels:} Sample~3.
All satellites located within a projected radius of
$r_p < 500$~kpc have been used in the calculations.
}
\label{fig2}
\end{figure}

The left panels of Figure~2 show $P(\phi)$ for the GIF satellites,
from which it is clear that they 
show a strong preference for alignment
with the major axes of the projected halo mass. The degree of anisotropy
in the satellite location angles is nearly identical for all three host--satellite
samples.
The right panels of Figure~2 show
$P(\phi \le \phi_{\rm max})$ for the location angles of the
satellites (solid lines), as well as $P(\phi \le \phi_{\rm max})$ for
the location
angles of the mass particles that are contained within the host halos
(dashed lines). 
From the right panel of Figure~2, then,
the distribution of satellite galaxies relative to the major axes of 
their hosts is very similar to the distribution of the mass
particles in the projected halos.
The satellites show a slightly flatter distribution than the mass particles,
but overall the satellites trace the projected shapes of the halos rather well.
We demonstrate this further in the top panel of Figure~3, where we show
the median location angle of the satellites and the mass particles as a function
of the ellipticity of the projected halo.  The median location angles
decrease approximately linearly with halo
ellipticity, although the slope is a bit steeper for the satellites than
it is for the mass particles.  The relationship is especially linear for
$0.2 < \epsilon_{\rm halo} < 0.35$, yielding slopes of
$-57^\circ \pm 2^\circ$ for the 
satellites and $-49^\circ \pm 0.9^\circ$ for the mass particles.

\begin{figure}
\centerline{
\scalebox{0.75}{%
\includegraphics{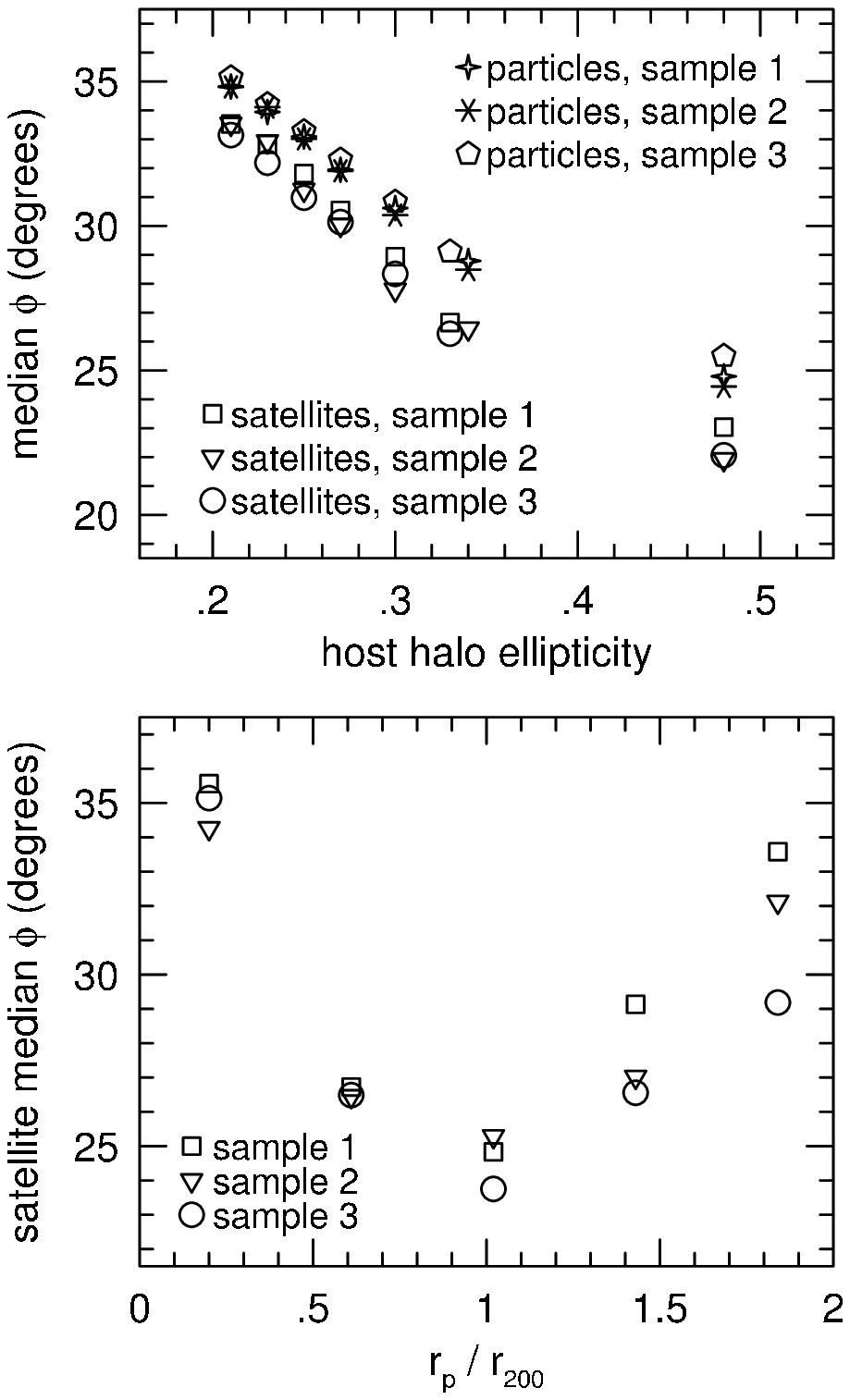}%
}
}
\caption{
{\it Top panel:} Median location angles of satellites and halo particles
as a function of the
ellipticity of the projected halo.  All satellites with $r_p <
500$~kpc and all particles within $r_{200}$ have been used in the calculation.
{\it Bottom panel:} Median location angles of satellites as a function of projected
radius on the sky, scaled by the virial radius of the host 
halo.  The median host halo virial radius is $\sim 275$~kpc in all three samples.
Bins have been chosen such that an equal number of
objects falls within each bin.
}
\label{fig3}
\end{figure}

Brainerd (2005) found that the anisotropy of the location angles of SDSS satellite
galaxies was most pronounced on small scales
($r_p \lesssim 100$~kpc) and that satellites with $r_p \gtrsim 100$~kpc
were distributed rather isotropically.  Brainerd (2005) speculated that
this could indicate that the virial region of the host halos
extended to only about 100~kpc, with satellites at larger radii being
part of an infalling population.
We investigate this possibility in the
bottom panel of Figure~3, where we show the median satellite location
angle as
a function of the satellite's projected radius, scaled by the virial
radius of the halo (i.e., $r_p/r_{200}$).  From this figure, then, the
anisotropy in the satellite location angles is present over all
scales, but it appears to be most
pronounced for satellites with $r_p \sim r_{200}$.  In addition, the anisotropy
persists to projected radii of order $2r_{200}$.  This result compares rather
poorly to the observation that the satellites of SDSS hosts seem to
be distributed isotropically on large scales, and the resolution of this
discrepancy is not immediately obvious.

Finally, to allay any lingering concerns that overmerging of the 
satellite population could affect the distribution of the location
angles of the GIF satellites, we compute $P(\phi)$ and 
$P(\phi \le \phi_{\rm max})$ for satellites that are present at $z=0$
but which are known to have been in existence since at least $z=0.52$.  That is,
we restrict our analysis to satellites in the file galsl\_sdss.cat\_1178
that can also be found in the file galsl\_sdss.cat\_0671, the $z=0.52$
galaxy catalog in which magnitudes are given in the 
SDSS band passes.  (Note that $z=0.52$ is the highest redshift for which
galaxy magnitudes are available in the SDSS band passes.)  
Roughly 60\% of the satellites that are present at $z=0$ can be traced
back to $z=0.52$ as unique objects.
Shown in
Figure~4 are $P(\phi)$ and $P(\phi \le \phi_{\rm max})$ for these satellites
that are known to be ``old'' (filled points and dotted lines, respectively)
compared to $P(\phi)$ and $P(\phi \le \phi_{\rm max})$ for all satellites
at $z=0$ (open points and dashed lines, respectively).  The probability
distributions at $z=0$ for the ``old'' satellites are statistically identical
to the probability distributions for all satellites at $z=0$, and we therefore 
conclude that the anisotropy in the satellite location angles at $z=0$ is 
independent of the ages of the satellites and is not strongly affected
by overmerging in the simulation.

\begin{figure}
\centerline{
\scalebox{0.65}{%
\includegraphics{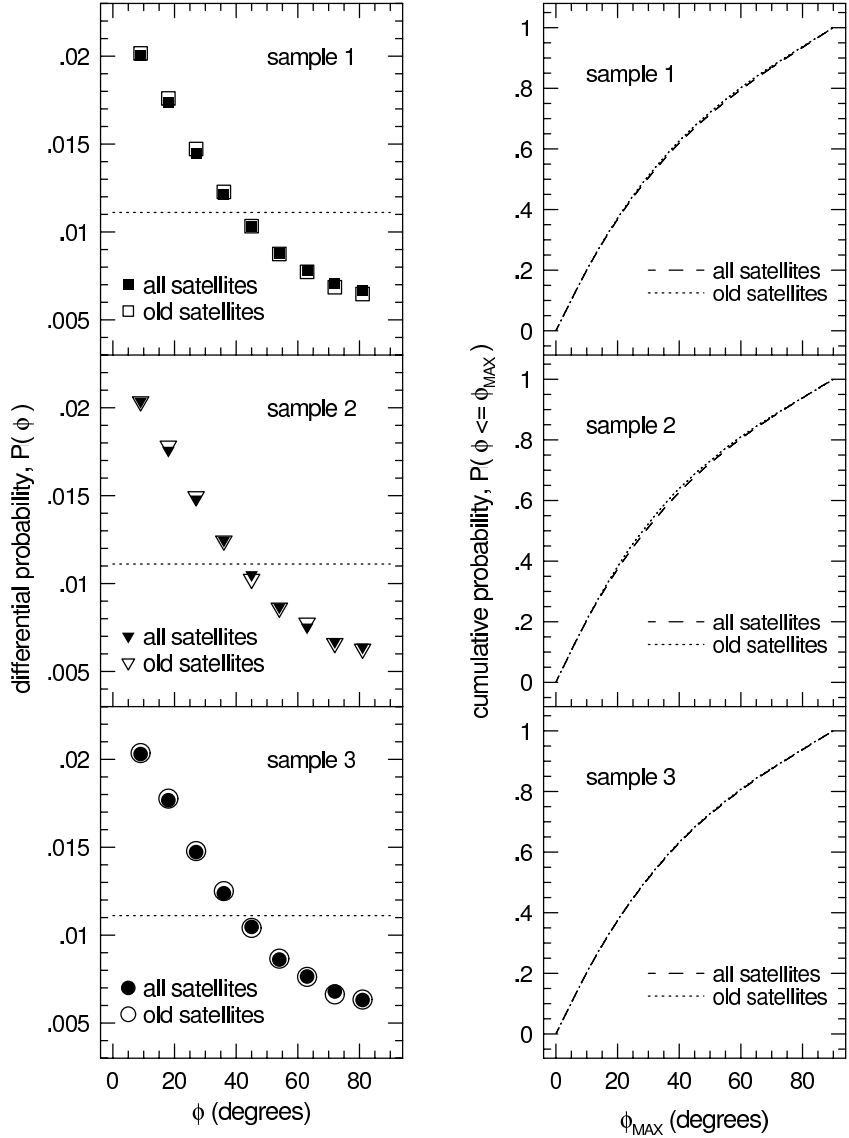}%
}
}
\caption{
{\it Left panels:} Differential probability distributions for
the location angles of satellites
at $z = 0$.  Open points show the results for all satellites, filled
points show the results for satellites that have existed as unique
objects since at least $z = 0.52$.  {\it Right panels:} Cumulative probability
distributions for the location angles of
satellites at $z=0$.  Dashed lines show the results for
all satellites, dotted lines show the results for satellites that have
existed as unique objects since at least $z=0.52$.
}
\label{fig4}
\end{figure}

\section{Misalignment of Light and Mass in the Hosts?}

The sense of the anisotropy that was found in
the previous section is in excellent agreement with the sense of
the anisotropy shown
by the SDSS satellites (i.e., a preference for location nearby the major
axes of the host galaxies).  However, the {\it size} of the effect is
grossly different.  The median location angle for the GIF satellites in
the previous section is $\phi_{\rm med}^{\rm GIF} \sim 29^\circ$, while for
the SDSS satellites the median location angle
is $\phi_{\rm med}^{\rm SDSS} \sim 40.5^\circ$ (e.g.,
Brainerd 2005).  In other words, if the major axes of the light emitted by
the GIF hosts were perfectly aligned with the major axes of their projected
halos, our results in \S 3 predict that the distribution of satellites about
their host galaxies should be much
more anisotropic than is observed in our universe.

There are at least two possible explanations for this.  First, in the previous
section we focused on halos with projected ellipticities of $\epsilon_{\rm halo}
 > 0.2$.  
This necessarily biases the sample of satellite galaxies in the simulation 
compared to observed galaxies in the universe. That is, since one cannot see
the dark matter halos of host galaxies in the real universe, one cannot
reject the satellites of host galaxies whose halos happen to be rather round
in projection.  We know from the numerical work of
Libeskind et al.\ (2005)
and Zentner et al.\ (2005) that in CDM universes the luminous satellites of
Milky Way--type halos tend to lie
in highly--flattened structures that are embedded in the principle
planes of the host halos.  This is a natural consequence of the infall of mass
and galaxies along filaments.
Thus, by choosing to use only host halos with substantial projected
ellipticities in the previous section, we have introduced a bias that 
optimizes the degree to which the satellite distribution is observed to
be flattened.

Further, if our previous assumption of perfect alignment of mass and light 
in the hosts is incorrect, this
would contribute to the excessive anisotropy 
shown by the GIF satellites.  We will
investigate this possibility in this section 
by creating artificial structures 
to represent the luminous regions of the host galaxies.  These artificial
structures will be embedded within the hosts' halos using a number
of different prescriptions, and the location angles of the satellite galaxies
will then be computed relative to the major axes of the images
that result from projecting the artificial structures
onto the sky.

We have visually inspected the images of the 200 brightest SDSS host galaxies in
each of Samples 1, 2, and 3 from Brainerd (2005) and we find that 95\% of
these objects are disk galaxies.  This is unsurprising since the host--satellite
selection algorithm yields only the very most isolated host galaxies, and large
ellipticals are known to be quite rare in low density environments.  Also, although
there are no images of the GIF hosts in the simulation, we can make a rough
assessment of whether the GIF hosts are more likely to be late--type or 
early--type galaxies based upon their $B$--band disk--to--bulge ratios.  That is,
early--type galaxies are 
expected to have $M(B)_{\rm bulge} - M(B)_{\rm total} < 1$~mag.\ 
(e.g., Simien \& de Vaucouleurs 1986).  
$B$--band bulge luminosities are not provided for the
GIF hosts and, therefore, we use the photometric transformation given by
Smith et al.\ (2002) to assign $B$--band magnitudes:
$$
B = g' + 0.47(g' - r') + 0.17
$$
and compute
$M(B)_{\rm bulge} - M(B)_{\rm total}$ for the GIF hosts.   
Unfortunately, bulge magnitudes are not reported for 35\% of our GIF hosts.
Of those hosts for which bulge magnitudes are reported, however,
only 13\% are consistent with being early--type galaxies.
Given the observed morphologies of the SDSS hosts and the relative
strengths of the bulges of the GIF hosts, then, it seems reasonable to adopt
a circular disk as the artificial structure that will represent the luminous
regions of the GIF hosts in this section.

In order to place the artificial circular disks within the host halos, we 
adopt four different methods for defining the orientations of
the disk angular momentum
vectors, $\vec{J}$: (1) $\vec{J}$ is aligned with the minor
(i.e., ``$c$'') axis of the halo's virial mass, (2) $\vec{J}$ is aligned with 
major (i.e., ``$a$'') axis of the halo's virial mass, (3) $\vec{J}$ is aligned
with the net angular momentum vector of the halo's virial mass, and (4) 
$\vec{J}$ is aligned with the intermediate (i.e., ``$b$'') axis of the mass
that is contained within a radius of 2.5~Mpc, centered on the host.  In the first
case we are making the na\"ive assumption that the disk lies in the principle
plane of the halo.  In the second case, we are assuming that the disk lies 
perpendicular to the halo's major axis.  This choice is the most sought after
solution for the Holmberg effect, despite the seemly unnatural orientation of
the disk relative to the halo (e.g., Libeskind et al.\ 2005; Zentner et al.\ 2005).
In the third case we are simply assuming that the angular momentum vectors of the
luminous and dark material are aligned.  Our fourth choice for the orientation of
the host disks is motivated by work by Navarro, Abadi \& Steinmetz (2004)
that showed the angular momentum vectors of disk galaxies in CDM universes
tend to align with the intermediate axes of the inertia tensor of the local
large--scale structure at turnaround.  Here we use the inertia tensor of the local
large--scale structure at the present epoch but, as Navarro, Abadi \& Steinmetz
(2004) note, the orientations of the principle axes of the inertia tensor do not
change substantially between turnaround and the present.  Throughout, we
refer to our four methods of placing the disks in the host halos as
$\vec{J}_c$, $\vec{J}_a$, $\vec{J}_{\rm ang}$, and
$\vec{J}_{\rm lss}$ respectively.

In order to insure that the major axes of the images of host galaxies are well--defined
in observational data sets, analyses of the location angles of satellite galaxies
with respect to the host major axes are generally restricted to hosts whose
images are clearly non--circular (e.g., Sales \& Lambas 2004; Brainerd 2005).
Following these observational studies, then, we restrict our analysis below
to those hosts whose disks have ellipticities $\epsilon_{\rm disk} > 0.2$ in
projection on the sky.  We use host--satellite selection criteria identical
to those in \S2,  and in Table 1 we list the mean number of hosts and 
satellites in a given rotation of the simulation.  Note that the luminosity
distributions, mass functions, 
distribution of the number of satellites per host, distribution of
apparent magnitude differences, and distribution of
stellar mass ratios (e.g., Figure~1) are not
affected by the increased number of hosts and satellites compared to \S3.

In this section, the major axis of a host galaxy's virial mass 
is not guaranteed
to be aligned with the major axis of the host's projected circular
disk. In \S3 we simply assumed that the offset, $\Delta \theta$,
between the major axis of the virial mass of the host galaxy and
the major axis of its image would be zero.  Shown in the left
panels of Figure~5, however, are the actual probability
distributions, $P(\Delta\theta)$, that we obtain when we place
circular disks within the halos. The form of $P(\Delta \theta)$ is
essentially unaffected by the way in which the hosts and
satellites are selected (e.g., Samples 1, 2, or 3 for a given choice of
disk orientation).  However,  $P(\Delta \theta)$ is strongly affected by our choice
of the orientation of disk angular momentum vectors.
In the case of the $\vec{J}_c$ hosts, the light is
fairly well aligned with the virial
mass in projection on the sky.  The median value of
$\Delta \theta$ is $6^\circ$ and only one third of the hosts have
values of $\Delta \theta$ that exceed $11^\circ$.
For the $\vec{J}_a$ hosts, the light is
essentially perpendicular to the virial mass.  Here
the median value
of $\Delta \theta$ is $87^\circ$ and only one third of the hosts have
values of $\Delta \theta$ less than $84.5^\circ$.
For the $\vec{J}_{\rm ang}$
hosts, the light is somewhat aligned with the virial mass,
the median value of $\Delta \theta$
is $\sim 32^\circ$ and one third of the hosts have values of
$\Delta \theta$ that exceed $\sim 48^\circ$. 
For the $\vec{J}_{\rm lss}$ hosts, the
orientation of the light is almost completely
unrelated to the orientation of the virial mass; 
the median value of $\Delta \theta$ is $\sim
41^\circ$, with one third of the hosts having a value of $\Delta
\theta$ that exceeds $\sim 56^\circ$.

\vskip 0.5cm 
\centerline{\bf Table 1: Mean Number of Hosts and
Satellites}
\medskip
\centerline{
\begin{tabular}{lrr}
 Sample & $N_{\rm hosts}$ & $N_{\rm sats}$ \\ \hline\hline
Sample 1, $\vec{J}_c$ & 2828 & 9417 \\
Sample 2, $\vec{J}_c$ & 480 & 1889 \\
Sample 3, $\vec{J}_c$ & 1510 & 4786 \\ \hline
Sample 1, $\vec{J}_a$ & 2934 & 9725 \\
Sample 2, $\vec{J}_a$ & 528 & 2064 \\
Sample 3, $\vec{J}_a$ & 1607 & 5065 \\ \hline
Sample 1, $\vec{J}_{\rm ang}$ & 2471 & 5809 \\
Sample 2, $\vec{J}_{\rm ang}$ & 443 & 1164 \\
Sample 3, $\vec{J}_{\rm ang}$ & 1345 & 3109 \\ \hline
Sample 1, $\vec{J}_{\rm lss}$ & 2864 & 9522 \\
Sample 2, $\vec{J}_{\rm lss}$ & 516 & 2021 \\
Sample 3, $\vec{J}_{\rm lss}$ & 1558 & 4905 \\ \hline\hline
\end{tabular}
}
\vskip 0.5cm

\begin{figure}
\centerline{
\scalebox{0.65}{%
\includegraphics{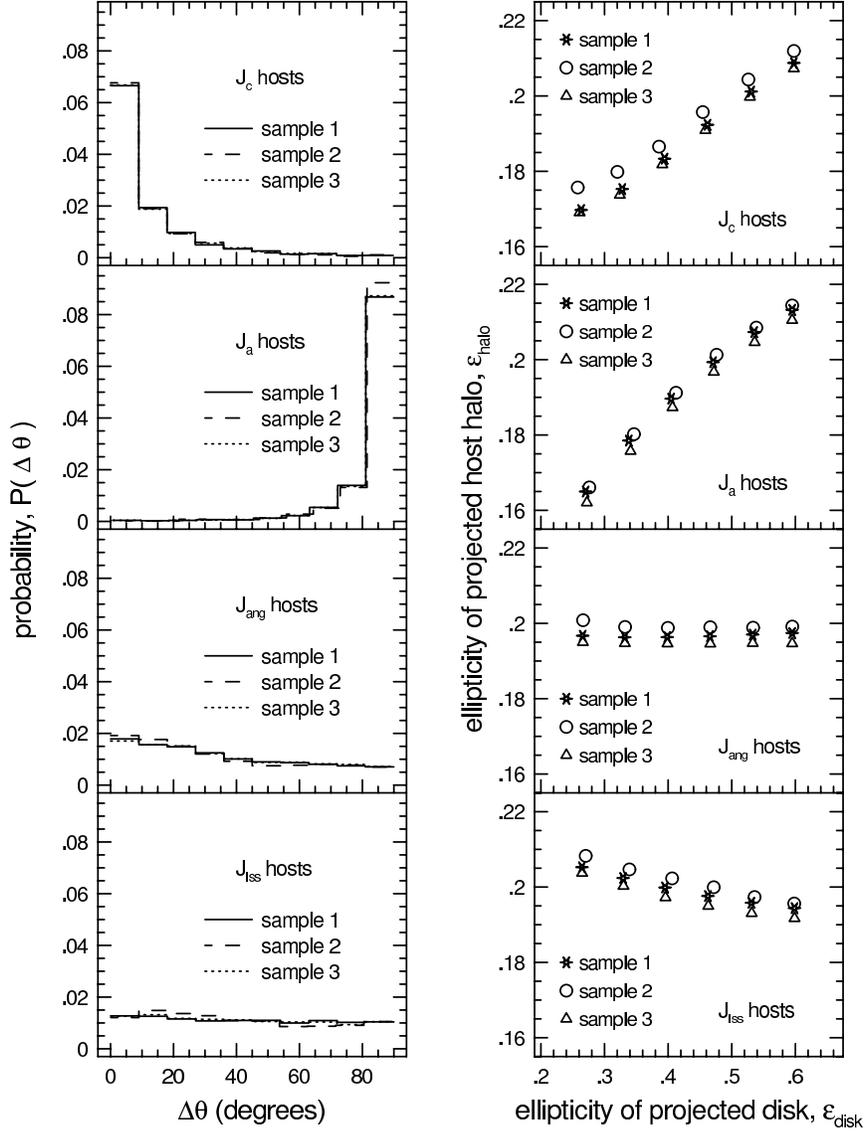}%
}
}
\caption{
{\it Left panels:} Probability distribution, $P(\Delta \theta)$, of the offset
between the major axis of a host galaxy's
projected halo mass 
and the major axis of the host's projected circular disk.
Different line types indicate Sample 1 (solid lines),
Sample 2 (dashed lines) and Sample 2 (dotted lines).
{\it Right panels:} Median ellipticity of the projected host halos as a
function of the ellipticity of the projected disks.  Different point
types indicate Sample 1 (stars), Sample 2 (circles) and Sample 3 (triangles).
}
\label{fig5}
\end{figure}

Shown in the right panels of Figure~5 are the median values of the
projected ellipticities of the host halos, $\epsilon_{\rm
halo}$, as a function of the ellipticities of the projected circular
disks, $\epsilon_{\rm disk}$. Overall, there is a very weak
dependence of the median value of $\epsilon_{\rm halo}$ on
$\epsilon_{\rm disk}$.  In other words, the selection of host galaxies on
the basis of a substantial flattening of their {\it images} is not
equivalent to selecting hosts on the basis of a substantial
flattening of their {\it halos}.  If, indeed, the locations of
the satellites of large
host galaxies in our universe are fairly good indicators of the 
existence of flattened dark matter halos,
this suggests that the mean anisotropy of the 
satellite location angles should be rather insensitive to the mean ellipticity of the
host images (i.e., $\left< \phi \right>$ should be only weakly dependent
on $\epsilon_{\rm disk}$).  To date, however, measurements of $\left< \phi
\right>$ for satellite galaxies as a function of the ellipticities of their
hosts has not been made with the available redshift surveys
(i.e., SDSS, 2dFGRS).  This is due primarily to the fact that the signal
to noise is only barely sufficient to detect the anisotropic distribution
of the satellites using the full data set; subdividing the data set into
bins of various values of $\epsilon_{\rm disk}$ simply reduces the signal
to noise to the point that no conclusive statement can be made about the
anisotropy.

In the case of the 
$\vec{J}_c$ and $\vec{J}_a$ GIF hosts, the median
value of $\epsilon_{\rm halo}$ increases with $\epsilon_{\rm
disk}$,  but the trend is not dramatic. Linear
extrapolation of the points in the top right panel of Figure~5 
yields median values of
$\epsilon_{\rm halo} \sim 0.26$  for edge--on, $\vec{J}_c$
host disks and
$\epsilon_{\rm halo} \sim 0.14$ for face--on,
$\vec{J}_c$ host disks.  Nearly identical results are obtained
for the $\vec{J}_a$ hosts.
In the case of the $\vec{J}_{\rm ang}$ hosts, the median
value of $\epsilon_{\rm halo}$ is  essentially independent of $\epsilon_{\rm
disk}$.
Finally, for the $\vec{J}_{\rm lss}$ hosts
the median value of $\epsilon_{\rm halo}$ is a decreasing function
of $\epsilon_{\rm disk}$, leading to very slightly rounder projected
halos for edge--on hosts disks and very slightly flatter projected halos for
face--on host disks.

The reason that the median projected halo ellipticity is
at best weakly dependent on the ellipticity of the host's projected
disk is that the
probability distributions of the halo ellipticities, $P(\epsilon_{\rm halo})$,
are only weakly dependent on $\epsilon_{\rm disk}$.  We illustrate
this in Figure~6 where we show $P(\epsilon_{\rm halo})$ for the
$\vec{J}_c$ hosts as a function of the ellipticities of the
projected disks.  Since
$P(\epsilon_{\rm halo})$ changes only mildly
from panel to panel in Figure~6, it is clear that
selecting host
galaxies on the basis of highly--elliptical images does not
preferentially select the very flattest halos, nor does it exclude
the very roundest halos.  This in mind, we expect the location angles
of the satellites of hosts that are selected on the
basis of the flatness of their images (as opposed to the flatness
of their dark matter halos) will show much less anisotropy than we
found in \S 3 above.

\begin{figure}
\centerline{
\scalebox{0.65}{%
\includegraphics{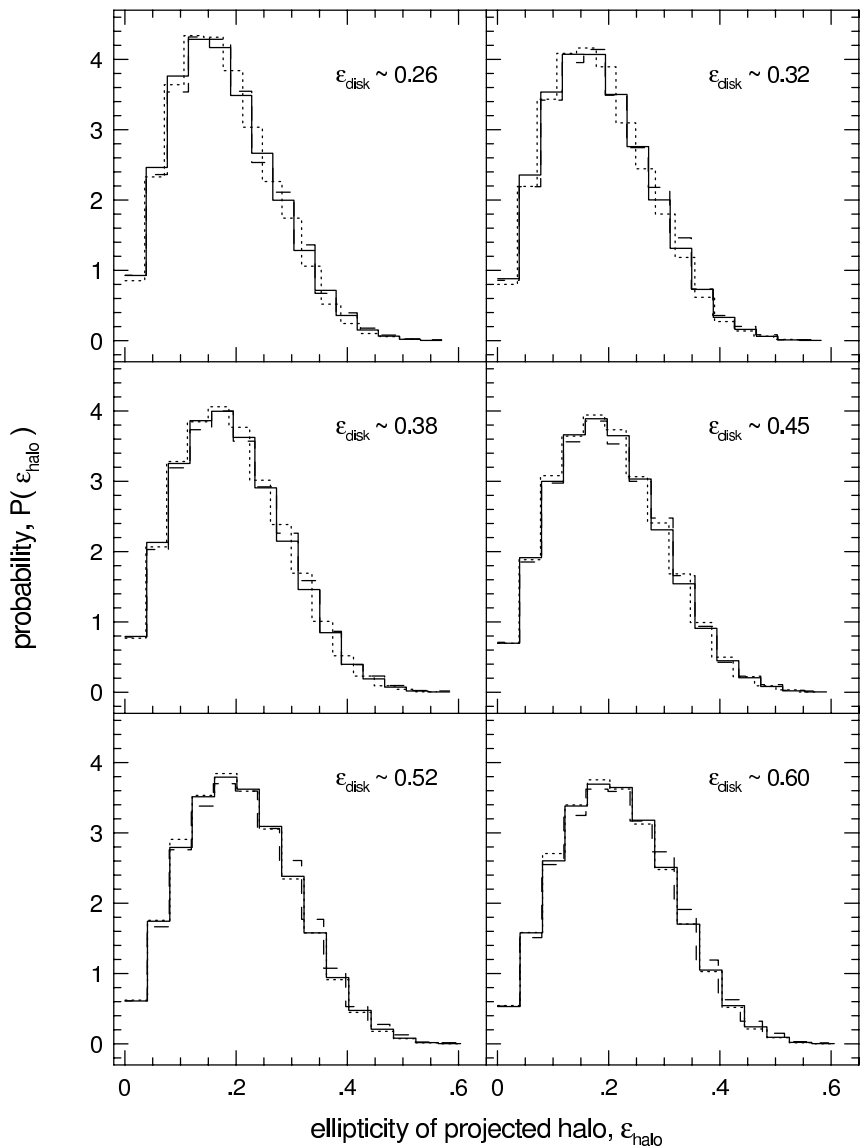}%
}
}
\vskip -1.5cm
\caption{
Probability distribution of the projected host halo ellipticity,
$P(\epsilon_{\rm halo})$, for $\vec{J}_c$ hosts with different
projected disk ellipticities, $\epsilon_{\rm disk}$.
Different line types show Sample 1 (solid lines), Sample 2 (dashed lines)
and Sample 3 (dotted lines).
Host halos are defined to be all particles within $r_{200}$.
}
\label{fig6}
\end{figure}

Like Figure~2 in the previous section, Figure~7 shows the differential
and cumulative probability distributions for the location angles of
the GIF satellites.  Here, however, $\phi$ is measured with
respect to the major axes of the hosts' projected circular disks 
rather than with respect to the major axes of the hosts' projected
halos.  This change in the definition of the symmetry axis that is
used to measure the location angles of the satellites has a marked affect
on the probability of a satellite having a given location
angle, $\phi$.
In the case of the $\vec{J}_c$ hosts, the hosts' circular disks
lie in the principle planes of the halos and the satellites clearly 
still 
prefer alignment with the major axes of the hosts.  However, the degree
of anisotropy is reduced from that in Figure~2 and here the 
median value of the location angle is $\phi_{\rm med} \sim 35^\circ$
(c.f.\ $\phi_{\rm med} \sim 29^\circ$ in the previous section).
Aligning the angular momentum vectors of the hosts' disks with the net
angular momentum vectors of the hosts' halos (i.e., the $\vec{J}_{\rm ang}$ hosts)
results in the satellites having
a rather weak preference for being aligned with the major axes of their
hosts ($\phi_{\rm med} \sim 42^\circ$).  When the angular momentum 
vectors of the hosts disks are aligned with the intermediate axes of 
the local large scale structure (i.e., the $\vec{J}_{\rm lss}$ hosts) the distribution
of satellite location angles becomes nearly
isotropic ($\phi_{\rm med} \sim 44^\circ$).  
As expected from Figure~5, the misalignment of mass and light in these host
galaxies, as well as the inclusion of halos that are round in projection,
results in a reduction of the anisotropy that was found when the satellite
location angles were measured relative to the projected major axes of markedly flattened
host halos (i.e., \S3).
Finally, an extremely strong
``Holmberg effect'' is produced when the angular momentum vectors of the
hosts' disks are aligned with the major axes of the halos' virial mass.
That is, the satellites of the $\vec{J}_a$ hosts show a strong preference
for being clustered near the minor axes of the hosts, yielding a median location
angle of $\phi_{\rm med} \sim 57^\circ$.

\begin{figure}
\centerline{
\scalebox{0.65}{%
\includegraphics{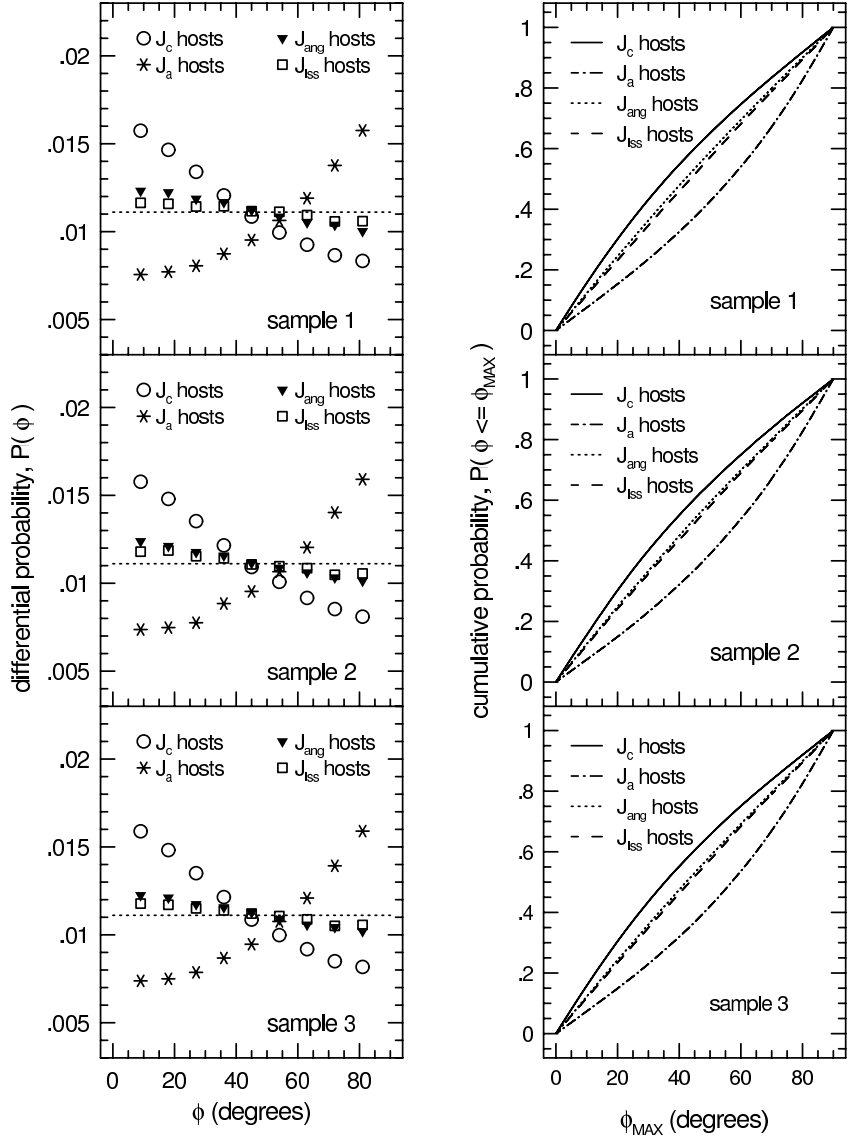}%
}
}
\caption{
Probability distributions for the location angles of satellite galaxies,
measured with
respect to the major axes of the projected disks of their hosts.  Hosts
are required to have a projected ellipticity of $\epsilon_{\rm disk} >
0.2$. 
All satellites with $r_p < 500$~kpc are used.
{\it Left panels:} Differential
probability distributions. 
Different point types show results for the $\vec{J}_c$ hosts (open circles), 
$\vec{J}_a$ hosts (stars),
$\vec{J}_{\rm ang}$ hosts (filled triangles),
and $\vec{J}_{\rm lss}$ hosts (open squares).
{\it Right panels:} Cumulative probability.  Different
line types show results for the
$\vec{J}_c$ hosts (solid lines),
$\vec{J}_a$ hosts (dash--dot lines),
$\vec{J}_{\rm ang}$ hosts (dotted lines),
and $\vec{J}_{\rm lss}$ hosts (dashed lines).
In all panels the axis scales are identical to the corresponding panels in
Figure~2 for comparison.
}
\label{fig7}
\end{figure}

Because the selection of hosts based upon flattened images (i.e., 
$\epsilon_{\rm disk} > 0.2$) is not equivalent to selecting hosts 
on the basis of flattened halos (i.e., Figures~5 and 6), we expect that
the degree of anisotropy shown by the satellite location
angles in this section should, at best,
be weakly dependent on the shape of the hosts' projected disks.  This is
shown by the left panels of Figure~8 in which the open points indicate
the dependence of $\phi_{\rm med}$ on the ellipticity of the hosts'
disks, $\epsilon_{\rm disk}$.  Also shown for comparison in these panels is
the dependence of $\phi_{\rm med}$ on $\epsilon_{\rm halo}$  from
the top panel of Figure~3 (crosses).  In the cases of the $\vec{J}_{\rm ang}$
and $\vec{J}_{\rm lss}$ hosts, the large offset in projected mass compared to
projected light results in $\phi_{\rm med}$ for the satellites being 
essentially independent of the ellipticity of the hosts' images.  In the
cases of the $\vec{J}_c$ and $\vec{J}_a$ halos, the degree of anisotropy
in the satellite location angles shows some dependence on $\epsilon_{\rm disk}$,
but it is not nearly as pronounced as the dependence of the anisotropy
on $\epsilon_{\rm halo}$.  For both the $\vec{J}_c$ and
$\vec{J}_a$ hosts, $\phi_{\rm med}$ for the edge--on
hosts differs from $\phi_{\rm med}$ for the hosts with $\epsilon_{\rm disk} \sim 0.2$
by of order $10^\circ$, with the anisotropy being most pronounced for the
edge--on hosts.

Finally, the dependence of $\phi_{\rm med}$ on projected radius, scaled
by the virial radii of the hosts, is shown in the right panels of
Figure~8.  As was the case for satellite location angles measured
relative to the major axis of the projected halo mass (i.e., bottom panel
of Figure~3), the degree of anisotropy shown by the satellites of the
$\vec{J}_a$ and $\vec{J}_c$ hosts is most pronounced for projected radii
that are of order $r_{200}$.  This dependence is not shown by
the satellites of the $\vec{J}_{\rm ang}$ and $\vec{J}_{\rm lss}$ hosts,
and in these cases $\phi_{\rm med}$ is largely independent
of projected radius.

\begin{figure}
\centerline{
\scalebox{0.65}{%
\includegraphics{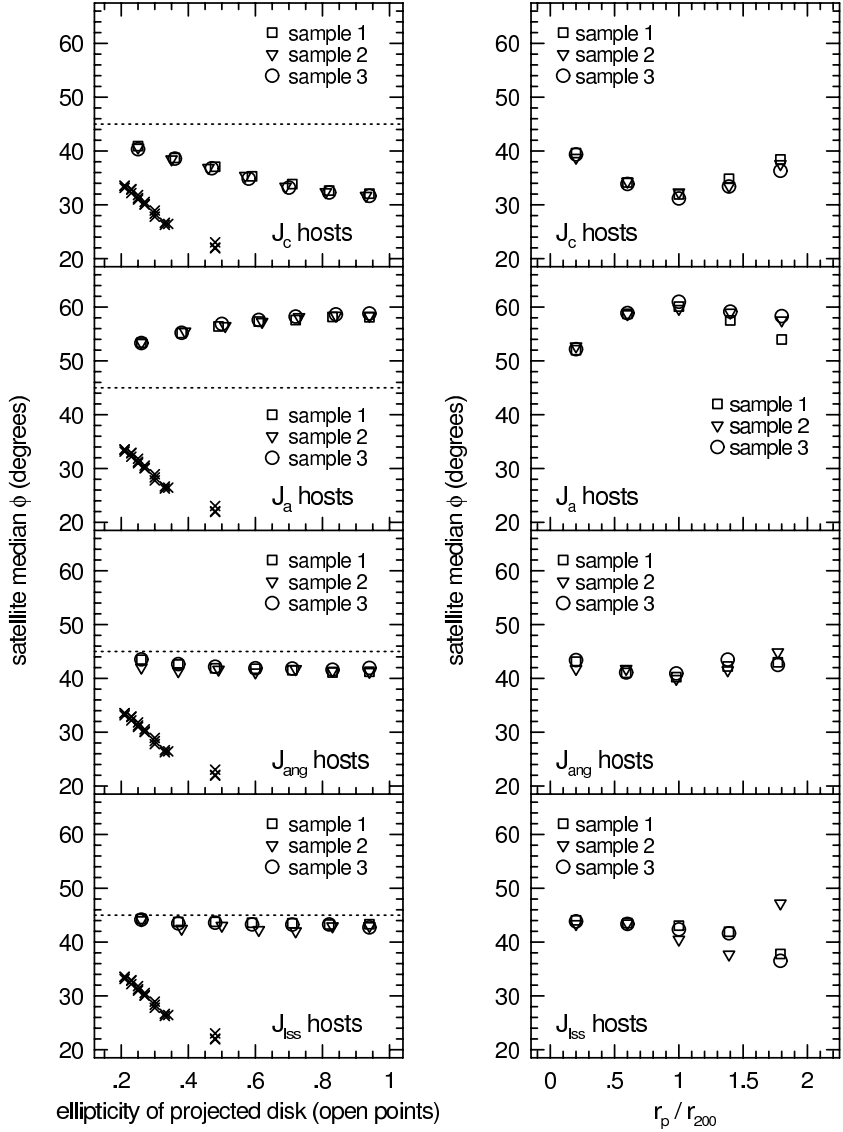}%
}
}
\caption{
{\it Left panels:} Open points show the median satellite location angle, measured
with respect to the major axis of the host galaxy's projected circular
disk, as a function of the ellipticity of the host's disk.
All satellites with $r_p < 500$~kpc have been used. For
comparison, crosses show the median satellite location angle from the top
panel of Figure~3 for all three samples as a function of the ellipticity
of the projected host halo (i.e., here $\phi$ is measured with
respect to the major axis of the projected host halo).   {\it Right panels:}
Median satellite location angle, measured with respect to the major axis
of the host galaxy's projected circular disk, as a function of projected
radius, $r_p$.  Here $r_p$ is measured in units of the host galaxy's
virial radius, $r_{200}$. 
}
\label{fig8}
\end{figure}

\section{Discussion}

We have used the $\Lambda$CDM GIF simulation
to investigate the locations of satellite galaxies with respect 
to the major axes of their hosts and find that:
\begin{enumerate}

\item When the location angles of the satellites are measured with respect to
the major axes of the projected host halos, the satellites exhibit an
anisotropic distribution that traces the projected halo mass 
rather well.  

\item If the mass and light of host galaxies are perfectly aligned in 
projection on the sky, then the sense of the anisotropy in the satellite
location angles is the same as that shown by satellite galaxies in the SDSS
(i.e., a preference for location nearby the major axis of the host's
image).  However, the magnitude of the anisotropy shown by the GIF satellites
far exceeds that of the SDSS satellites since our analysis was restricted
to the satellites of hosts with markedly flattened halos ($\epsilon_{\rm halo} 
> 0.2$).

\item If all host galaxies are disk galaxies, there is only a weak correlation
between the ellipticity of the projected disk, $\epsilon_{\rm disk}$, and the 
ellipticity of the projected halo, $\epsilon_{\rm halo}$.  No matter how
the disk is oriented within the halo, this alone reduces the anisotropy 
in the satellite location angles.
This is because the satellites essentially trace the 
projected mass of the hosts' halos,
and round halos are not rejected by simply requiring the
image of the host to be rather elliptical (i.e., $\epsilon_{\rm disk} > 0.2$).

\item If the location angles of the satellite galaxies are measured relative to the
major axes of their host's projected disks, the resulting degree of anisotropy 
is strongly dependent upon the way in which the
host disk is oriented inside its halo.  A pronounced ``Holmberg effect'' (i.e.,
preference for location nearby the minor axes of the hosts) is
obtained when the angular momentum vectors of the host disks are 
aligned with the major principle axes of the hosts' halos.  A strong tendency
for the satellites to be found nearby the major axes of the hosts (i.e.,
the type of anisotropy shown by the SDSS satellites) is obtained when the angular
momentum vectors of the host disks are 
aligned with the minor principle axes of their halos.  When the angular
momentum vectors of the host disks are aligned with the net
angular momentum vectors of their halos, the satellites have a tendency to
be located nearby the major axes of their hosts, but the degree of anisotropy
is considerably weaker than when the angular momentum vectors  of the
host disks are aligned with the minor axes of their halos.
When the angular momentum vectors of
the host disks are aligned with the intermediate principle axes of the
local large scale structure, the location angles of the satellites become nearly
isotropic.

\end{enumerate}

We have shown that, provided mass and light are ``reasonably'' well--aligned
in the host galaxies, $\Lambda$CDM naturally predicts that satellite galaxies
should should be anisotropically distributed relative to the symmetry axes
of their hosts and, specifically, 
the satellites should show a preference for being located
nearby the major axes of their hosts.  The sense of this anisotropy 
agrees well with the anisotropy shown by the satellites of host galaxies in the 
SDSS, but none of our simple prescriptions for defining the major axes of
host galaxies in the GIF simulation precisely reproduces the magnitude of
the anisotropy shown by the SDSS satellites ($\phi_{\rm med}^{\rm SDSS} \sim 
40.5^\circ$).  When we assume all host galaxies are disks and we align the
disk angular momentum vectors with the minor principle axes of the host
halos, the GIF satellites are distributed more anisotropically than are
the SDSS satellites ($\phi_{\rm med}^{\rm GIF} \sim 35^\circ$).  If, instead,
we align the host disk angular momentum vectors with the net angular momentum
vectors of their halos, the GIF satellites are distributed less anisotropically
than are the SDSS satellites ($\phi_{\rm med}^{\rm GIF} \sim 42^\circ$).  In
addition, the anisotropy shown by the GIF satellites appears to persist to
much larger projected radii than does the anisotropy shown by the SDSS
satellites.

It is important to keep in mind that the GIF and SDSS hosts are by no means
identical, if for no other reason than the resolution limit of the simulation
restricts our analysis to rather massive isolated host galaxies.  In
addition, since there are no images of the hosts in the simulation, we have
used very simple arguments to precisely align the angular momentum
vectors of the GIF hosts with their surrounding environments.  In particular,
we have had to use large physical scales (i.e., from $r_{200}$ to 
2.5~Mpc) to define the surrounding environment.  However, work by
Bailin et al.\ (2005) on the formation of disk galaxies in CDM universes 
suggests that the orientation of the disk tends to be aligned with only
the inner halo, and that the relative orientation of the inner and outer
halo are essentially uncorrelated.  Further, at least some fraction of
the SDSS hosts are elliptical galaxies and we have not allowed for this in our
analysis.  Finally, many of the host galaxies in the SDSS are obvious
spirals and an exact determination of the ``major axis'' of such
hosts is difficult because the light profiles are not smooth.  This naturally
introduces some level of error in the observational determination of the
locations of satellite galaxies around their hosts, and serves to decrease
the measured anisotropy compared to the true anisotropy that one would
obtain if the host position angles were known to arbitrary accuracy.

There are two previous numerical studies of the locations of 
luminous satellite galaxies that are directly comparable to our work here.
Both Libeskind et al.\ (2005) and Zentner et al.\ (2005) used semi--analytic
galaxy formation to study the locations of
luminous satellites (i.e., rather than extrapolating
from a pure dark matter simulation).  
There are, however, a number of important differences between the Libeskind et al.\
(2005) and Zentner et al.\ (2005) investigations and the investigation
presented here. First, both Libeskind et al.\ (2005) and
Zentner et al.\ (2005) used simulations with much smaller volumes and
much higher resolution
in order to address
the apparent planar clustering of the Milky Way's satellite galaxies.
In their simulations
Libeskind et al.\ (2005) and Zentner et al.\ (2005) were
restricted to the
study of only a small number of hosts (6 and 3, respectively), each of which
had a large number of satellites, and the distribution
of the satellites was computed independently for each host halo.  Here we have
used a large number of hosts with a wide range of masses and luminosities and,
since most of our hosts have only a few satellites, our results for the distribution
of the satellites is obtained by effectively ``stacking'' all of
the host--satellite systems together.  That is, Libeskind et al.\ (2005)
and Zentner et al.\ (2005) investigated what $\Lambda$CDM
predicts specifically for host galaxies that are like the Milky Way, while our
study investigates what $\Lambda$CDM predicts for large redshift surveys in which one
is limited to only a few satellites per host.

Neither Libeskind et al.\ (2005) nor Zentner et al.\
(2005) selected host--satellite systems in a way that mimics what is commonly
done with large redshift surveys.  Libeskind et al.\ (2005) combined semi--analytic
galaxy formation with their N--body simulations and, so, there are truly
luminous satellite galaxies (as opposed to simply ``subhalos'') present in the
simulations.  The satellites used by Libeskind et al.\ (2005) consist of the 11
most luminous satellites of each host galaxy, in analogy with the Milky Way system.
Zentner et al.\ (2005) use two techniques to select satellites:
semi--analytic galaxy formation and an association of satellites with the most
massive surviving subhalos.
Both Libeskind et al.\ (2005) and
Zentner et al.\ (2005) conclude that the satellite populations of their
Milky Way systems are strongly flattened
into thin, somewhat disk--like structures that are aligned with the longest
principle
axis of the dark matter halos of the host galaxies.  They further conclude that,
in the case of the Milky Way, the observed distribution of the satellites implies
that the longest principle axis of the Milky Way's halo is oriented perpendicular
to the disk. (It should be noted, however, that there are no actual luminous disks
for the host galaxies in these simulations so neither study demonstrated directly
that the disks of host galaxies are anti--aligned with the longest principle
axes of the halos.)

Our work here agrees well with the sense of the anisotropy of the satellite locations
found by Libeskind et al.\ (2005) and Zentner et al.\ (2005); the satellites are
found in a flattened, rather than spherical, distribution and that distribution
is aligned well with the longest principle axis of the halo of the host galaxy.
In addition, we concur with their result that satellite galaxies should be located
preferentially close to the minor axes of the images of their hosts if
the host galaxy is a disk that has its angular momentum aligned with the 
major principle axis of its halo (i.e., the satellites of our $\vec{J}_a$ hosts).
However,
since there are typically only one or two GIF satellites per host galaxy, we
cannot address the question as to whether the satellites of any one host
galaxy are generally found to lie in an extremely flattened, nearly-planar structure.
Like the results of Libeskind et al.\ (2005) and Zentner et al.\ (2005), 
however,
our results for the anisotropic distribution of the GIF satellites certainly
do seem to
be a reflection of the fact that satellites are accreted preferentially along
filaments.

Whether or not the locations of satellite galaxies in CDM universes
agree with the locations of satellite galaxies in large redshift surveys
remains an open question.  Here we have shown that the sense of the
observed anisotropy (a preference for clustering of satellites nearby
the major axes of the images of host galaxies) is consistent with the
sense of the anisotropy that one would expect in a CDM universe, under the
assumption that the major axes of the images of host galaxies are at least
modestly correlated with the major axes of their projected halos.  A proper
resolution to the discrepancies that we find between observation
and theory awaits simulations that are of considerably higher resolution
and which address the details of the orientations of the visible hosts
with respect to their dark matter halos.

\section*{Acknowledgments}

We grateful to Mario Abadi, Brad Gibson, and Julio Navarro for
discussing their work with us in advance of its publication and 
to Simon White for helpful suggestions.
We also thank the
referee for truly constructive criticisms that improved the manuscript.
Support under NSF contract AST-0406844 (IA, TGB) is
gratefully acknowledged.

\end{document}